\documentstyle[prl,aps]{revtex}
\begin{document}
\draft

\twocolumn[\hsize\textwidth\columnwidth\hsize\csname@twocolumnfalse\endcsname

\title{Measuring equilibrium properties in aging systems}
\author{Silvio Franz\cite{Silvio}}
\address{Abdus Salam International Center for Theoretical Physics\\
Strada Costiera 11, P.O. Box 563,
I-34100 Trieste (Italy)}
\author{Marc M\'ezard\cite{Marc}}
\address{Laboratoire de Physique Th\'eorique de l'Ecole
Normale Sup\'{e}rieure \thanks{Unit\'e propre du CNRS,  associ\'ee
 \`a\ l'Ecole
 Normale Sup\'erieure et \`a\ l'Universit\'e de Paris Sud}\\
24 rue
 Lhomond, F-75231 Paris Cedex 05, (France)}
\author{Giorgio Parisi\cite{Giorgio}}
\address{Dipartimento di Fisica and Sezione INFN,\\
Universit\`a di Roma ``La Sapienza'',
Piazzale Aldo Moro 2,
I-00185 Rome (Italy)}
\author{Luca Peliti\cite{Luca}}
\address{Dipartimento di Scienze Fisiche and Unit\`a INFM\\
Universit\`a ``Federico II", Mostra d'Oltremare, Pad.~19, 
I-80125 Napoli (Italy)}
\date{\today}

\maketitle

\begin{abstract}
We corroborate the idea of a close
connection between replica symmetry breaking and aging in the linear
response function
for a large class of finite-dimensional systems with short-range interactions.
In these system, characterized by a continuity condition with respect
to weak random perturbations of the Hamiltonian, the ``fluctuation
dissipation ratio'' in off-equilibrium dynamics should be equal to
the static cumulative distribution function of the overlaps.
This allows for an experimental measurement of the equilibrium order
parameter function.
\end{abstract}

\twocolumn\vskip.5pc]\narrowtext

The glassy state of matter can appear in systems with quenched disorder
(like spin-glasses), or in non-disordered systems. Ergodicity breaking 
takes a special form in these systems. A rather generic  
situation is the existence of many
solid, ``glass'', phases, which are very different from one another, 
and unrelated among themselves by
symmetry transformations. 
Hence the Gibbs equilibrium measure decomposes into
a mixture of many pure states. This
phenomenon was first studied in detail in the
mean field theory of spin glasses, where it received the name of 
replica-symmetry breaking 
\cite{SGtheo}. But it can be defined in a straightforward way and easily
extended to other systems, by considering an 
order parameter function,
the overlap distribution function. This function measures 
the probability that two configurations of
the system, picked up independently with the 
Gibbs measure, lie at a given distance from each other \cite{par83}. 
Replica-symmetry breaking
is made manifest when this function is nontrivial.

The existence of non trivial overlap distributions, first found 
in mean field systems, has been shown unambiguously, 
through numerical simulations, in 
finite-dimensional spin-glass systems with short-range interactions 
 \cite{romani-young}.  
 This order parameter function is a very important
tool for the mathematical description of the Gibbs state. 
Unfortunately it seems impossible
to access it experimentally for two reasons: (1) Large glassy systems never 
reach equilibrium at low temperatures; (2) The measurement of the 
distance between configurations requires 
a detailed observation at the microscopic---atomic---level, which is impossible.
(In simulations, the second objection disappears, and one can 
get around the
first one by working with smart algorithms and small enough systems.) 

The first objection is a very basic one: 
experimentally, glassy systems exhibit a non-equilibrium behavior,
which requires a dynamical description. 
Quite often, they exhibit a special type of dynamical
behavior called aging, i.e.,
the property that extensive one-time quantities like the energy,
magnetization etc.,
are  asymptotically close to  time-independent values, 
whereas two-time quantities, like the
autocorrelation functions and their associated linear response functions,
continue to depend on the time elapsed after the quench even for long
times.
Aging, defined in this way,
appears in mean-field spin glasses and has been
exhibited in spin-glass experiments \cite{age-exp,francesi}. 
We will not discuss systems undergoing 
``stabilization'' \cite{zarzycki} (sometimes called ``physical aging''
\cite{struik}), 
where one-time quantities cannot be considered close to their asymptotic 
values during typical experiments. (This phenomenon has been recently observed 
in a lattice gas model with constrained dynamics \cite{KPS}.)
In aging dynamics the usual equilibrium properties do not hold. 
The analysis of some spin-glass mean-field models \cite{cuku}
has suggested in particular that
the usual fluctuation-dissipation relation between the correlation and the
response should be modified in a well-defined
way. This modification, which holds when both the age of the system
and the measurement time are large, involves the rescaling of the temperature
by a
``fluctuation dissipation ratio" ({\sc fdr}), 
which depends on the relation between
the two times involved \cite{FM,cukusk}. This {\sc fdr} can be found 
experimentally
 by simultaneous measurements of the noise and the
response on various time scales and age scales. 

The aim of this paper is twofold. We shall first show that, in {\it finite
dimensional\/} systems with {\it short range\/} interactions, 
there exists an identity relating
the---experimentally accessible---{\sc fdr} 
to an equilibrium order parameter function.
This static order parameter function
is an interesting new object. 
We shall then discuss its relationship to the usual 
distribution of  overlaps.
 Our argument relies on
a perturbation of the original hamiltonian 
by the addition of some weak---but thermodynamic---random perturbations.
This method has been recently used to derive interesting properties
of the overlap distribution at equilibrium \cite{guerra,aizenmann-contucci}.

We use the language of magnetic systems, and
denote by $S_x$ the spin at a point 
$x$ of a lattice of size $L^d$ in $d$ dimensions. 
We work with classical spins which are
real variables in a double well potential, 
and the Ising limit will often be considered for
simplicity. We call $H(S)$ the Hamiltonian. 
Our argument is rather general and we do not
have to specify much the Hamiltonian: 
it contains short range interactions, in a $d$
dimensional space; it may contain quenched disorder or not. 
The evolution of the spin dynamics is
governed by the Langevin equation
\begin{equation}
\dot{S_x}=-{\partial H\over\partial S_x}
+\eta_x ,
\end{equation} 
where $ \eta_x$ is a white noise of variance 
$\langle \eta_x(t) \eta_y(t')\rangle 
= 2 T \delta_{xy} \delta(t-t')$. 
(We denote by angular brackets thermal averages,
i.e., either, in the dynamic framework, the average 
with respect to the realization of 
the random noise, or, in the static framework, 
the average with respect to the Gibbs
measure). 
The system starts at time $t=0$ from a random initial condition.
Important quantities are the correlation function, 
$C(t,t')= (1/N) \sum_x \langle S_x(t) S_x(t')\rangle$, 
and the response function, which measures the response
of the spins at time $t$ to an instantaneous field at time $t'$:
\begin{equation}
R(t,t')={1\over N}\sum_x 
\frac{\delta \langle S_x(t) \rangle}{\delta \eta_x(t')}  .
\end{equation}
The quantity which is measured experimentally (thermoremanent magnetization)
 is the integrated response function, defined  by
$
\chi(t,t')= T \int_0^{t'} d t'' R(t,t'') 
$.

The {\sc fdr} $X(q)$ is  
obtained by considering the infinite-time limit
of the response function, fixing 
the correlation function $C(t,t')$ to a given value $q$ \cite{cuku,FM,cukusk}:
\begin{equation}
X(q)=\lim_{{t,t'\to\infty}\atop{C(t,t')=q}}
\frac{\partial \chi(t,t') }{\partial t'}\Big/
\frac{\partial C(t,t') }{\partial t'}.
\label{fdrat}
\end{equation}

The usual equilibrium dynamics is obtained by sending the two times $t,t'$ to
infinity while keeping their difference $\tau=t-t'$ fixed. Then the correlation
and response functions,
$C(t,t')$ and $R(t,t')$ reach their equilibrium values, $c(\tau)$ and $r(\tau)$. 
 In short-range systems
this regime relates to the property of ``local equilibrium'', i.e., to the
fact that any finite region of space reaches equilibrium
locally. The Edwards-Anderson
order parameter is defined dynamically by 
$q_{EA}=\lim_{\tau \to \infty} c(\tau)$,
and the usual fluctuation-dissipation theorem
asserts that, for $q>q_{EA}$, $X(q)=1$. 
The aging regime concerns systems
with weak ergodicity breaking, such that the correlation 
$C(t,t')$ relaxes below $q_{EA}$
when $t \to \infty$ (at fixed $t'$) \cite{bouchaud}. 
Then the {\sc fdr} $X(q)$ can become different
from unity in the regime $q<q_{EA}$. A simple way to measure it is through 
a parametric plot of the integrated response function 
versus the correlation \cite{francesi}.
The ratio $T/X(q)$  can be interpreted as an effective temperature \cite{teff}.

We wish to relate the {\sc fdr} to an equilibrium order parameter.
Let us add to the original Hamiltonian a 
 perturbation of the form $\epsilon H_2$, with
\begin{equation}
H_2=\sum_x h_x S_x S_{T(x)},
\end{equation}
where the $h_x$'s are independent gaussian random variables of variance one,
and $T$ is a translation of length $L/2$ in a fixed direction $e$, 
say the $x$ axis (so that $T(x)=x+(L/2)e$).
The thermal expectation value of the perturbation
$
\langle H_2 \rangle
$,
is a contribution to the internal energy of the system which is extensive
and self-averaging, i.e., independent (in the thermodynamical limit)
of the particular realization of the disorder contained in either
$H$ or $H_2$.
 The interaction $H_2$,
which looks long-range,
is in fact a local  perturbation in a different space.
Let us divide the space into two halves (${\cal S}_{l}$ and 
${\cal S}_r$)
and rename the spins in the right hand part so that
 if $x \in {\cal S}_l$ then $T(x) \in {\cal S}_r$ and $S_{T(x)}=S'_x$. 
The total Hamiltonian can now be written 
as
\begin{equation}
H(S,S')=H_l({ S})+H_r({ S'})+B({ S,S'})
+\epsilon\sum_{x\in {\cal S}_l}
h_x S_x S_x'.
\label{hloc}
\end{equation}
The hamiltonians $H_l$ and $H_r$ refer 
respectively to the spins in ${\cal S}_{l}$ and 
${\cal S}_r$. The  term $B({ S,S'})$ is a surface term whose presence  does not
affect the average of $H_2$. Dropping it, the Hamiltonian (\ref{hloc})
characterizes a spin system of size $L^{d}/2$, with two spins
 $S_x,S_x'$ on each site, and a purely local interaction.
Notice that in the 
 case of disordered systems the spin systems $S$ and $S'$ taken individually
 contain two independent realizations of the disorder.
 
Since the perturbation $H_2$ is a sum of local terms, the thermal expectation
value (for almost all realizations of the disorder)
$
\langle H_2(t) \rangle
$
measured in the dynamics has a long-time limit which is equal to its
equilibrium expectation value. The proof of this fact is standard for
systems with short-range interactions. We first notice 
 that the free energy density $f(t)$ must
reach, at long times, its equilibrium value $f_{eq}$: 
if it were to converge to a value $f(\infty)$ larger than the equilibrium one, 
one could always nucleate
a bubble of radius $r$ with the equilibrium free energy, with
a cost of free energy less than or equal to $c r^{d-1} + (f_{eq}-f(\infty)) 
r^d$,
which becomes negative for large enough $r$. Therefore the free energy reaches
equilibrium, as well as its derivative with respect to $\epsilon$, proving
the convergence of $\langle H_2(t) \rangle$. (Notice that we do not discuss here
the time scale for reaching this equilibrium, which may become very long in some 
systems:
what matters here is that it is finite when $L \to \infty$.)

We  now compute the expectation value of the perturbation  $H_2$
in the dynamics  and in the statics. Since $\langle H_2(t) \rangle$
is self-averaging, it is equal to its average over the
random field $h$ and all other possible 
quenched disorder in the system, 
which we denote by $E_h\,\langle H_2(t) \rangle$.
In the dynamical framework, starting from the Langevin equation 
in presence of the perturbation 
$\epsilon H_2$, we express
the average of $H_2$ in the Martin-Siggia-Rose formalism 
\cite{ZJ} as a path integral:
\begin{eqnarray}
 \langle H_2 (t) \rangle & = & E_h\, \langle H_2(t)\rangle\nonumber\\
& = &
E_h  \int {\cal D}(S) \, {\cal D}(\hat{S})\; 
{\rm e}^{I[S,i\hat{S}]}
 {
\sum_x h_x S_{x}(t)
S_{T(x)}(t)} ,
\end{eqnarray}
with the dynamical action
\begin{equation}
I[S,i\hat{S}]= \int d t'\; \sum_x i\hat{S}_x(t') \left[ \dot{S_x}+
{\partial H\over\partial S_x}+\epsilon {\partial H_2\over\partial S_x}
 + iT\hat{S}_x\right].
\end{equation}
Integrating by parts over the $h_x$'s, and observing that the insertion 
of $i\hat{S}_x(t')$ acts as the derivative with respect to an impulsive
magnetic field at site $x$ and at time $t'$: $ \delta/\delta h_x(t')$, 
we obtain \cite{cukudyn}
\begin{equation}
 E_h\, \langle H_2(t)\rangle=2\epsilon\sum_x
E_h 
\frac{\delta}{\delta h_{T(x)}(t)}\langle S_x(t) S_x(t') S_{T(x)}(t')\rangle.
\end{equation}
In the linear response regime $\beta \epsilon\ll 1$, the average of 
the product 
on far away sites factorizes up to terms of order 
$\epsilon$ and one has 
\begin{equation}
E_h \frac{\delta}{\delta h_{T(x)}(t)}\langle S_x(t) S_x(t') S_{T(x)}(t')\rangle
=C(t,t') R(t,t') +O(\epsilon)
\end{equation}
Assuming that the bound holds uniformly in time 
(remember that the large volume limit is taken before the large time limit)
and substituting the definition (\ref{fdrat}) of the {\sc fdr}
we obtain for large values of $t$
\begin{equation}
2 \epsilon\beta N \int_{0}^1 dq X_\epsilon(q)  q= \epsilon \beta N\left( 1
-\int_{0}^1 dq \; {dX_\epsilon \over dq} \; q^2\right),
\label{edyn}
\end{equation}
where we have assumed  
$\lim_{t \to \infty} C(t,0)=0$ for simplicity. 
We have denoted by $X_\epsilon$ the {\sc fdr}  of the system
with the perturbed Hamiltonian.  
Notice that this is a very general result that holds for every
sample in the case where there is quenched disorder. 

We now turn to the statics. The thermal equilibrium average  of $H_2$
 is self-averaging with respect to disorder. We can thus
 evaluate it as follows:
\begin{equation}
\langle H_2\rangle
 = E_h{1\over Z} \sum_{S}
e^{-\beta\left[H( S)+\epsilon H_2( S)\right]} \sum_x
h_x S_{x}
S_{T(x)}.
\end{equation}
Integrating by parts over the $h_x$'s, we obtain
 \begin{equation}
\langle H_2 \rangle 
= \beta \epsilon N E_h \left(1- {1 \over N} \sum_{x}\langle  S_{x} S_{T(x)}
 \rangle^2 \right) 
\end{equation}
Invoking again linear response for small $\epsilon$, together with the fact that 
$x$ and $T(x)$ are infinitely far apart in the thermodynamic limit, we can write 
\begin{equation}
E_h \langle  S_{x} S_{T(x)}
 \rangle^2= E_h \langle  S_{x} S_{y}
 \rangle^2 +O(\epsilon) 
\end{equation} 
where $x$ and $y$ are two far away spins not directly coupled in $H_2$.
We obtain then (up to higher orders in $\epsilon$):
\begin{equation}
\langle H_2 \rangle 
= \beta\epsilon N E_h
\left(1-\int\,d q \;P_\epsilon(q)\; q^2\right) .
\label{estat}
\end{equation}
The last equality, involving the overlap distribution 
$P_\epsilon(q)$ for the perturbed system,
results from the decomposition of the Gibbs measure 
into a sum of pure states
characterized by a clustering property \cite{par83}.

Comparing the two results, (\ref{edyn}) and (\ref{estat}),
for the dynamics and the statics,
we see that the second moments of the dynamical order parameter
function $dX_\epsilon(q)/dq$ and of the static one $P_\epsilon(q)$ 
coincide for the 
system in presence of the perturbation $\epsilon H_2$. It is straightforward
to generalize this derivation to perturbations of the type 
$H_p=\sum_x h_x S_x S_{T_1(x)} \cdots S_{T_{p-1}(x)}$, where 
$T_k(x)=x +  (k/p L) e$. 
(For $p=1$ the perturbation is nothing but a small random
field term.) This shows that the $p$-th moments of the 
two functions $dX_\epsilon(q)/dq$  and $P_\epsilon(q)$ coincide.

Let us now consider the functions $ \tilde X(q)= 
\lim_{\epsilon \to 0} X_\epsilon(q)$ and 
$\tilde P(q) =\lim_{\epsilon \to 0}P_\epsilon(q)$. (To be precise, we need to
introduce simultaneously all the perturbations with arbitrary $p$ and
strength $\epsilon_p$, and send all the $\epsilon_p$'s to $0$.) 
These are two characteristic
functions of our problem. One describes the violation
of the {\sc fdt} in the out of equilibrium dynamics, and the other describes
some equilibrium correlations. These two functions are equal, and thus
an unexpected link between statics and dynamics is established.

We now discuss the relationship
between the new functions $d \tilde X/dq, \tilde P(q)$ and the more
conventional definitions of the {\sc fdr} and the overlap distribution.
 Let us first consider the
equilibrium distribution $\tilde P(q)$. Clearly, in a situation with ergodicity 
breaking and
several nearly degenerate pure states, the effect of the $\epsilon$ 
perturbation which scales
as $\epsilon \sqrt{L^d}$ induces a reshuffling of the weights of the states. 
A simple
example appears when there is an exact degeneracy due to a symmetry. 
For instance, in the case
of an Ising Hamiltonian quadratic in the spin variables, the
symmetry by reversal of all the spins implies that 
the overlap distribution $P(q)$ 
of the unperturbed system is
symmetric: $P(q)=P(-q)$, each pure state appearing with the same weight as
its symmetric one in the Gibbs measure. This symmetry will be violated by 
the perturbation terms
$H_p$ with odd $p$, leading (in the absence of further reshuffling) 
to the relation
$\lim_{\epsilon\to 0^+} P_\epsilon(q)\equiv P_{0^+} (q)= 2 \theta(q) P(q)$. 
Suppose now
that care has been taken of all the symmetries of the system, by defining
a modified $\hat P(q)$ measuring the distance between orbits of the
symmetry group.
 It is reasonable
to believe that, for a large class of systems, the reshuffling of the 
weights will only lift the degeneracy, so that $\tilde P(q)=\hat P(q)$. 
We shall
refer to such systems as being {\it stochastically stable}. 
Mean-field spin glasses 
fall into
this category, as well as Ising ferromagnets in
 dimensions $d \ge 2$\cite{fmpp},
but we do not know how to characterize this class in general. 
Turning to the case of dynamics, it is trivial to show that
the limit of $\epsilon \to 0$ is smooth when it is taken
before the limit of large times (the infinite volume
limit is always taken first). If the limits commute,
then the static $\tilde X(q)=\int_0^qdq'\,\tilde P(q')$ 
is identical to the dynamical ({\sc fdr}) $X(q)$ of
the unperturbed system, measured with {\it random\/}
initial spin configurations.
This results holds for Sherrington-Kirkpatrick
spin glasses, but is violated, e.g., in $p$-spin  spherical
spin-glass models, where the dynamics is dominated
by infinitely long-lived metastable states.

Summarizing, we have introduced a new order parameter function
for  systems at equilibrium, which can be related in general, in
finite dimensional systems, to the
{\sc fdr} of a weakly perturbed system. This new order parameter
is interesting since its moments are obtained as expectation values of
extensive quantitites. It is thus much more robust than the usual overlap
distribution, and will not have the same chaotic behaviour under weak 
perturbations. It is interesting to notice that in mean field spin glasses,
the order parameter appearing from the replica computation is 
naturally related to this new order parameter. 
For stochastically stable systems,
this new order parameter is 
equal to the 
overlap distribution of symmetry classes
of the states, and can be measured experimentally.
The relation
implies that in any finite-dimensional system 
replica-symmetry breaking and
aging in the response functions either appear together, or do not
appear at all. 
Although we have used the language of the magnetic systems, 
the arguments put forward here 
can be generalized to systems of different nature, using different 
random  perturbations of the Hamiltonian.

\acknowledgments
S.F. thanks Prof.~D. Sherrington for kind
hospitality at the Departement of Theoretical Physics of the
University of Oxford, where part of this work was elaborated.
We thank A. Barrat, A. Cavagna, I. Giardina,
D. Sherrington and 
M. Virasoro for valuable discussions.


\begin{thebibliography}{99}
\bibitem{cukudyn} A similar technique was first used in
\cite{cuku} in order to relate the {\sc fdr} to the expectation value
of a long range perturbation in the $p$-spin systems.
\end{thebibliography}
\end{document}